\documentclass[12pt]{article}
\usepackage{geometry}                		
\usepackage{graphicx}				
\usepackage{amsmath}								
\usepackage{amssymb}
\usepackage{amscd}
\usepackage{yfonts}
\usepackage{amsmath,amsthm,amssymb}
\usepackage{amsthm}
\newtheorem{theorem}{Theorem}

\newtheorem{definition}[theorem]{Definition}

              \newtheorem{lemma}[theorem]{Lemma}

\newcommand {\cH}{{\cal H}}

\newcommand {\bp} {{\bf p}}

\newcommand {{\bx}} {{\bf x}}
\newcommand {{\bk}} {{\bf k}}
\newcommand {\bP} {{\bf  P}}

\newcommand {\bR} {{\mathbb{R}}}
\newcommand  {\cA} {\mathcal{A}}

\date{}

\medskip

\title{Scattering  in algebraic  approach to quantum theory. Associative algebras}
\author{Albert Schwarz}
\date{}							

\begin{document}
\author {  A. Schwarz\\ Department of Mathematics\\ 
University of 
California \\ Davis, CA 95616, USA,\\ schwarz @math.ucdavis.edu}


\maketitle
\abstract
{The definitions of scattering matrix and inclusive scattering matrix in the framework of formulation of quantum field theory in terms of associative algebras with involution are  presented.
The scattering matrix is expressed in terms of Green functions on shell (LSZ formula) and the inclusive scattering matrix is expressed in terms of generalized Green functions on shell. 
The expression for inclusive scattering matrix can be used also for quasi-particles (for elementary excitations of any translation-invariant stationary state, for example, for elementary excitations of equilibrium state.) An interesting novelty is the consideration of associative algebras over real numbers.}
\section {Introduction}
The standard algebraic approach to quantum theory is based on consideration of associative algebra with involution $^*$(algebra of observables). This algebra will be denoted by $\cal A$, we assume that it has a unit. Usually it is assumed that $\cal A$ is an algebra over complex numbers (then the involution should be antilinear), but we  consider also the case when $\cal A$ is an algebra over real numbers.  Notice, however,  that   the case  of algebras over  real numbers is less natural; the reader can skip everything related to this case.

The states are defined as positive linear functionals on $\cal A$
(one says that the functional  $\omega$ is positive if $\omega (A^*A)\geq 0$ for $A\in \cal A$). The set of states is a convex cone denoted by $\cal C$. Proportional states are identified, hence  instead of the cone $\cal C$ one can consider the convex set  ${\cal C}_0$ of  states
obeying $\omega(1)=1$ (the set of normalized states).

Time translation (evolution operator) acts on $\cal A$ as involution preserving automorphism. To define particles and their scattering we need also spatial translations; together time and spatial translations span commutative group $\cal T.$ The time translations are denoted by $T_{\tau}$  and spatial translations by $T_{\bx}$ where ${\bx}\in \mathbb{R}^d$. The induced transformations of the space of states are denoted by the same symbols.  If $A\in \cal A$ we use the notation $A(\tau,\bx)$ for $T_{\tau} T_{\bx}A.$

Let us consider a stationary translation-invariant state $\omega$ and the pre Hilbert space $\cal H$ of the corresponding GNS (Gelfand-Naimark-Segal) representation of $\cal A$. Recall that  in this representation there exists a cyclic vector $\Phi$  obeying  $\omega (A)=\langle A \Phi,\Phi \rangle$. (One says that $\Phi$ is a cyclic vector if ${\cal H}={\cal A}\Phi$).  The translations descend to $\cal H$. In complex case the infinitesimal translations  (energy and momentum operators) are defined on a dense subset of  Hilbert space $\bar {\cal H}$ (of the completion of $\cal H$), in real case they act on a dense subset of the complexification of $\bar {\cal H}.$ (We use the same notation for translations in $\cal H$ as for automorphisms of $\cal A$. The  element $A$ of the algebra $\cal A$ and  the corresponding operator in $\cal H$ are also denoted by the same symbol.
Notice that  representing $A$ as an operator we should consider translation of $A$ as a conjugation with  $T_{\tau} T_{\bx}$, i.e.  $A(\tau,{\bx})=T_{\tau} T_{\bx}AT_{-\tau}T_{-\bx}.$ )

Elements of $\cal H$ can be regarded as excitations of $\omega.$

The {\it elementary space} $\textgoth {h}$ is defined as a 
space of smooth fast decreasing functions on $\mathbb{R}^d\times \cal I$  (all of their derivatives should decrease  faster than any power); it is equipped   with $L^2$ metric. (Here $\cal I$ denotes a finite set consisting of $m$ elements.)  We should consider real -valued  functions if $\cal A$ is an algebra over $\mathbb {R}$ and  complex-valued functions if $\cal A$ is an algebra over $\mathbb {C}.$ It is convenient to consider the elements of $\textgoth {h}$ as  columns of $m$ functions on $\mathbb{R}^d.$ The spatial translations act naturally on this space; we assume that the time translations also act on $\textgoth {h}$ and commute with spatial translations. In momentum representation the spatial translation $T_{\bx}$ is represented as multiplication by $e^{i\bx\bk}$ and the time translation $T_{\tau}$ is represented as a multiplication by a matrix $e^{-i\tau E(\bk)}.${\it We assume that $E(\bk)$ is a non-degenerate Hermitian matrix.  }  

To guarantee that time translations act in $\textgoth {h}$ we assume that $E(\bk)$ is a smooth function of $\bk$ and has at most polynomial growth. If $\textgoth {h}$ consists of complex-valued functions then diagonalizing the matrix $E(\bk)$ we can reduce  the general case to the case when $m=1$; this remark was used in \cite {SC}. ( Notice, however, that the eigenvalues of  $E(\bk)$ are not necessarily smooth functions of $\bk$.)

An  {\it elementary excitation} of $\omega$ is defined as an  isometric map  $\Phi : \textgoth {h}\to {\cal H}={\cal A}\Phi$ commuting with translations.

If $\cal A$ is an algebra over $\mathbb {R}$ we assume that the elements of $\textgoth {h}$ in momentum representation obey the reality condition $f^*({\bk})=f(-\bk).$

We show that imposing the condition of asymptotic commutativity we can define the scattering matrix  and inclusive scattering matrix of elementary excitations of $\omega.$ To analyze the properties of scattering matrix we assume that $\omega$ has cluster property.
 Our results are based on ideas of  \cite {SC} . ( This paper was  published as Section 13.3  of \cite {MO}, an improved version of it was published recently in preprint form.)  Notice that  the results of  \cite {SC} generalize  Haag-Ruelle  theory dealing with scattering matrix  in Lorentz-invariant local quantum field theories. (See \cite {AH} for  an exposition of Haag-Ruelle theory closest  to our approach and \cite {FS}  or \cite {MO} for generalization of this theory to the case when Lorentz-invariance is not assumed.)

If  $\cal A$ is a $C^*$-algebra over 
complex numbers (as in \cite {SC})  one can identify  (quasi-local) observables with self-adjoint elements of this algebra. For every normalized state $\omega$ one defines a probability distribution of the   observable $a$ corresponding to a self-adjoint element $A$ in such a way that for every continuous function $f$  the expectation value of $f(a)$ is equal to $\omega (f(A)).$  One can consider also  global observables corresponding to infinitesimal automorphisms of $\cal A
$, in particular, energy and momentum corresponding to time and spatial infinitesimal translations.  We can talk about joint spectrum of energy and momentum  operators  in $\cal H$ (in the space of excitations of translation-invariant stationary state $\omega$). We say that $\omega$ is a ground state if  the energy operator in $\bar {\cal H}$ is positive definite.

For algebras over real numbers one should consider skew-adjoint elements ($A=-A^*$) instead of self-adjoint elements (for algebras over complex numbers there exists an obvious one-to-one correspondence $A\to iA$ between skew- adjoint and self-adjoint elements). The definition of the probability distribution of physical quantity used in the case of complex numbers does  not work, however, one can use the geometric approach to quantum theory to derive the probability distribution from decoherence \cite {GA}.




 Elementary excitations of ground state are called particles. Elementary excitations of arbitrary translation-invariant stationary state are called quasi-particles. Quasi-particles are in general unstable; this means that the conditions in the definition of elementary excitation are satisfied only approximately. For quasi-particles the definition of scattering matrix does not work, however, the definition of inclusive scattering matrix still makes sense.

Notice that  our considerations can be applied to the scattering of elementary excitations of any translation-invariant state.  In particular, they can be applied to the scattering of elementary excitations of an equilibrium state. It is important to notice that our considerations can be applied also in non-equilibrium situation. The generalized Green functions we are using coincide with functions considered in Keldysh formalism of non-equilibrium statistical physics. They appear also in the formalism of $L$-functionals that can be used to give a simple and transparent derivation of  diagram technique for calculation of generalized Green functions in the framework of perturbation theory \cite {T},\cite {S},\cite {MO}.

The ground state is not singled out 
in any way in our considerations. 

The main goal of present paper is to give an exposition of scattering theory in such a way that it can be easily compared with the definition of scattering in geometric approach to quantum field theory \cite {GA3} and in the approach based on Jordan algebras \cite {GA4}.  
In the present
 paper we consider also  the case when $\cal A$ is an algebra over $\mathbb {R}$; this  is necessary for comparison with  papers \cite {GA3}, \cite {GA4} where we consider both real and complex elementary spaces. There exists an opinion that complex numbers are important in the formulation of quantum mechanics. It is true that imposing very natural axioms 
 one can  justify this opinion (see, for example,\cite {KA}).  One of  goals of \cite {GA} and present paper  is to formulate axioms that allow us to avoid using complex numbers.

\section {Scattering matrix. }
Let us consider an algebra with involution $\cal A$. We assume that spatial and time translations act as automorphisms of $\cal A.$  We fix a translation-invariant  stationary state $\omega$; excitations of $\omega$ are defined as elements of pre Hilbert space $\cal H$ obtained by GNS construction.  The algebra $\cal A$ acts in $\cal H$. In what follows we denote the operator in $\cal H$   corresponding to an element $A\in \cal A$  by the same letter. We assume that all operators we are dealing with are smooth. 
 (An operator corresponding to an element  $B\in \cA$ is smooth if $B=\int \alpha (\bx,t)  A(\bx,t)d\bx dt$, where $\alpha (\bx,t)\in {\cal S}(\bR^{d+1}), A\in \cA.$)

The element of $\cal H$ corresponding to $\omega$ is denoted by $\Phi.$ We define  an elementary  excitation as an isometric map of an elementary space $\textgoth {h}$ into $\cal H$  commuting  with spatial and time translations. (Recall that an elementary space consists  of smooth fast decreasing functions depending on spatial variable
$\bx$ or on momentum variable $\bk$ and on discrete variable taking $m$ values. We consider simultaneously algebras over complex an real numbers;
correspondingly the functions considered below are  complex-valued or real-valued.)

 Let us fix  $m$ elements $\phi_1,\cdots,\phi_m\in \textgoth {h}$ and $m$ operators $B_1,\cdots, B_m\in\cal A$ obeying $\Phi (\phi_i)=B_i\Phi.$  The elements $\phi_i$ are columns of functions $\phi_i^{\alpha}$, together they can be considered as a square matrix. {\it We assume that this matrix is invertible and commutes with the matrix $E(\bk)$ governing the time translation in $\textgoth {h}$.  If this condition is satisfied we say that  $B_1,\cdots, B_m$ are good operators. }
 We require also that   $B^*_i\Phi=\Phi(\psi_i)$ for some $\psi_i\in \textgoth {h}.$
 
 Notice that 
 $$ B_j(\tau,{\bx})\Phi =\Phi (T_{\tau}T_{\bx}\phi _j).$$ 
In momentum representation 
$$B_j(\tau,{\bx})\Phi=\Phi ({(e^{-i\tau  E{(\bk)}})}^{\alpha}_{\beta}e^{i\bk\bx}\phi_j^{\beta}).$$

Let us consider a collection of  smooth  functions $f=(f^1({\bk}),\cdots,f^m({\bk}))$ decreasing faster than any power.  ( The space of smooth fast decreasing functions is denoted by $\cal S$; we can say that $f\in {\cal S}^m.$ ) We define an operator $B(f,\tau)$
acting in $\overline {\cal H}$ (in the completion of $\cal H$) by the formula
\begin {equation}
\label {B}
B(f,\tau)= \int d{\bx} \tilde f ^j_{\tau}( {\bx})B_j(\tau,\bx)
\end {equation}
where $ \tilde f ^j_{\tau}( {\bx})$ denotes the inverse Fourier transform with respect to $\bk$ of the function 
$f^j_{\tau}({\bk})=f^i{(\bk)}(e^{i\tau  E{(\bk)}})^j_i.$

The operator $B(f,\tau)$ depends linearly of $f$; it specifies a  generalized vector function  of ${\bk},\tau$ that can be written in the form 
\begin{equation} \label {BB}
B({\bk},\tau)=e^{{i\tau  E{(\bk)}})}\hat B(\tau,\bk)
\end{equation}
where $\hat B(\tau,\bk)$ is a Fourier transform with respect to $\bx$ of $B_j(\tau,\bx)$ considered as a generalized vector function.

Let us prove that $B(f,\tau)\Phi$ does not depend on $\tau.$ Using ({\ref {B}}) we obtain
$$ B(f,\tau)\Phi=\int d{\bx}\tilde f^i_{\tau}({\bx})\Phi ({(e^{-i\tau  E{(\bk)}})}^{\alpha}_{\beta}e^{i\bk\bx}\phi_i^{\beta})=$$
\begin {equation} \label {BBB}\Phi (f^j{(\bk)}(e^{i\tau  E{(\bk)}})^i_j{(e^{-i\tau  E{(\bk)}})}^{\alpha}_{\beta}\phi_i^{\beta})=\Phi (f^j\phi_j^{\alpha}).
\end {equation}
(We have used the fact that the matrix $e^{i\tau  E{(\bk)}}$ commutes with the matrix $\phi$.)

In what follows we denote $f^j\phi_j^{\alpha}$ as $f\phi$ where $f$ is considered as column vector and $\phi$ as a square matrix.

Later we will use this statement in the following form:
\begin {lemma}\label {DD}
$\dot B(f,\tau)\Phi=0$
\end {lemma}
Here dot denotes the derivative with respect to $\tau.$
\begin {definition}\label {KK}
Let us consider the function $f^j_{\tau}({\bx})$ corresponding to the collection $f=(f^1,...,f^m)$ of smooth fast decreasing functions.
We say that a set  $\tau K(f)$ is an essential support  of the function $f^j_{\tau}({\bx})$ if for all $n$ 
$$f^j_{\tau}({\bx})<  C_n (1+|{\bx}|^{2}+\tau^2)^{-n}$$
where $ \frac {{\bx}}{\tau}\notin K.$
\end {definition}

In the case  when Fourier transforms $f^i(\bk)$ of functions $f^i(\bx)$ have compact support  one can assume that $K(f)$ is compact, but in general is not clear that one can find a compact set  $K $ obeying  the conditions of this definition.

Let us impose the conditions of asymptotic commutativity on the operators $ A_j\in \cA.$ This means that
\begin{equation}\label {AS}||[A_j(\tau, {\bx}),  A_k]||< \frac {C_n(\tau)}{1+||{\bx}||^n}.
\end {equation}
  Here $n$ is an arbitrary integer and $C_n (\tau)$ is a polynomial. ( The condition we have imposed can be weakened, see \cite {SC}.)



Let us consider the vectors
$$\Psi (f_1,\tau_1, ..., f_n,\tau_n)= B(f_1, \tau_1) ...B(f_n,\tau_n)\Phi$$
where $f_i\in {\cal S}^m$ is a collection of  $m$ smooth fast decreasing functions on $\mathbb {R}^d.$

We say that   $f_1,..., f_n$ are not-overlapping if the sets  ${K_j}={K(f_j)}$  do not overlap (more precisely we should assume that the distance between sets $K_j$ and $K_{j'}$ is positive for $j\neq j'$).

\begin {lemma}\label {NO}
If   $f_1,...,f_n$ do not overlap the vectors 
$$\Psi (f_1,\tau_1, ..., f_n,\tau_n)= B(f_1, \tau_1) ...B(f_n,\tau_n)\Phi$$ have a limit  in $\bar {\cal H}$ as $\tau_j$ tend to $-\infty$; this limit will be denoted by 
$$\Psi (f_1, ..., f_n| -\infty)$$
The set spanned by such limits will be denoted by $\cal {D}_-.$
 \end {lemma}
 Let us sketch the proof of this lemma  for the case when $\tau_1=...=\tau _n=\tau.$ It is sufficient to check that $\int _{-\infty}^0||\dot \Psi (\tau)||d\tau$ is finite. (Here $\Psi(\tau)$ stands for
 $\Psi (f_1,\tau, ..., f_n,\tau)$ and $\dot \Psi$ denotes the derivative with respect to $\tau$.) The derivative  $\dot\Psi$ is a sum of $n$ terms; every term contains $n$ factors and one of this factors is a derivative. We can interchange the factors because the commutators   can be neglected as $\tau \to \-\infty$; this follows from the condition that the functions $f_j$ do not overlap and from the fact that for  non-overlapping families  essential supports of functions $f^i_{\tau}({\bx},j) $ and $f^i_{\tau}({\bx},j')$   are far away for large $\tau$. 
 
We use this remark to shift the factor with the derivative to the right. It remains to apply the lemma \ref {DD}. 
 
  Let us define the $in$-operators $a^+_{in} $ by the formula
 \begin{equation}\label {IN}
a^+_{in}(f\phi)= a^+_{in}(f ^j\phi_j^{\alpha})= \lim_{\tau\to -\infty} B (f,\tau).
\end {equation}
 Lemma \ref {NO} gives  conditions on $f$ that guarantee the existence of  this limit  as strong limit on the set $\cal {D}_-.$ 

Let us introduce the asymptotic bosonic Fock space $\mathcal {H}_{as}$ as a Fock representation of 
canonical  commutation relations
$$[b(\rho), b(\rho')]=[b^+(\rho),b^+(\rho')]=0, [b(\rho),b^+(\rho')]=\langle \rho,\rho'\rangle$$
where $\rho, \rho'\in \textgoth {h}.$

We define M\o ller matrix $S_-$ as a map $\mathcal {H}_{as}\to \overline {\cal H}$ obeying
$$a_{in}^+(\rho) S_-=S_-b^+(\rho), S_-|0\rangle=|0\rangle$$
Here $|0\rangle$ stands for the Fock vacuum.

Notice that spatial and time translations act naturally in   $\mathcal {H}_{as}.$  The M\o ller matrix 
commutes with these translations.

Operators $a_{in}(\rho)$ are defined on the image of $S_-$ by the formula 
$$a_{in}(\rho) S_-=S_-b(\rho).$$
They are Hermitian conjugate to $a_{in}^+(\rho).$

One can give a direct definition of M\o ller matrix by the formula 
$$S_-b^+(f_1^k\phi_k^{\alpha})\cdots b^+(f_n^k\phi_k^{\alpha})|0\rangle=\Psi (f_1, ..., f_n| -\infty)$$
or equivalently
$$S_-(b^+(g_1)\cdots b^+(g_n)|0\rangle)=\lim _{\tau\to-\infty}B(g_1\phi^{-1},\tau)\cdots B(g_n\phi^{-1},\tau)\Phi.$$

If $g_1,\cdots, g_n$ do not overlap the vector
\begin{equation}\label {VE}
B(g_1\phi^{-1},\tau)\cdots B(g_n\phi^{-1},\tau)\Phi
\end {equation}
describes $n$ distant particles as $\tau\to -\infty.$

It is convenient to require strong  cluster property ( see Section 2 of \cite {SC}) to analyze the M\o ller matrix . (This condition can be weakened.)




Let us make the following
\begin {quote}\label {N}
{\bf Assumption}.
{\it The subset of the space ${\cal S}^{mn}$ spanned by non-overlapping families $(f_1,...,f_n)$  contains an open dense subset of}
${\cal S}^{mn}.$
\end{quote}

This assumption is not restrictive; see the discussion in Section 4.2 of \cite {SC}.

Using the above assumption  and cluster property one can prove  the theorem below.
\begin{theorem} \label {MMM}
The M\o ller matrix $S_ -$ is a well defined isometric operator.
\end {theorem}
Notice first of all that it is not clear from our definitions that the $in$-operators and M\o ller matrices are well defined (they can depend on the choice of operators $B_j$).  In other words, the operator $S_-$ a priori can be multivalued.  However, we can prove that this operator is isometric and use the fact that an isometric operator is necessarily single-valued. 

To  prove that the operator $S_-$ preserves the inner product
we express the inner product
$$\langle  B(f_1, \tau_1) ... B(f_n,\tau_n)\Phi,  B'(f'_1, \tau'_1) ...B'(f'_{n'},\tau'_{n'})\Phi\rangle$$
in terms of truncated correlation functions. We assume that both $(f_1,...,f_n)$ and $ (f'_1,...,f'_{n'})$ do not overlap, then it follows from Definition
 \ref {KK} and cluster property that only two-point  correlations $\langle B^{'*}(f',\tau')B(f, \tau)\Phi,\Phi\rangle=\langle B(f,\tau)\Phi, B'(f',\tau')\Phi\rangle$ contribute   in the limit $\tau_j,\tau'_{j'}\to -\infty.$ Calculating the two-point correlation functions by means of 
(\ref {BBB})  we see that  $S_-$ is an isometry.  We assumed that the vectors corresponding to families of non-overlapping functions span a dense subset of $\mathcal {H}_{as}$, hence $S_-$ can be extended to an isometric embedding of the  space $\overline {\mathcal {H}_{as}}$  into $\overline {\cal H}.$  

Taking $\tau\to +\infty$ instead of $\tau \to -\infty$ we obtain the definition of M\o ller matrix $S_+$  and of $out$-operators $a^+_{out}(\rho), a_{out}(\rho).$ If M\o ller matrices are  surjective operators we can define the scattering matrix ($S$-matrix)  by the formula $S=S_+^{-1}S_-$.
In this case one says that {\it the theory has particle interpretation.}\footnote { It is possible that the "elementary space" $\textgoth {h}$ does not describe all particles existing in the theory (for example, we are missing some composite particles). In this case we have a chance to get a theory with particle interpretation extending the space $\textgoth {h}.$}

In other words we can say that  the theory has particle interpretation if for dense subset of $\cal H$ the time evolution can be represented as linear combination of vectors
$$ e^{-i H\tau} \Psi (f_1, ..., f_n| -\infty)= \Psi (T_{\tau}f_1, ...,T_{\tau} f_n| -\infty)$$
for $\tau<0$
and of vectors
$$ e^{-i H\tau} \Psi (f_1, ..., f_n| +\infty)= \Psi (T_{\tau}f_1, ...,T_{\tau} f_n| +\infty)$$
fot $\tau>0.$

This means that generically both for $\tau \to -\infty$ and for $\tau\to +\infty$ we obtain a collection of distant particles (the wave functions $T_{\tau}f_k$ have distant essential 
supports if functions $f_k$ do not overlap).
Without assumption that the theory has particle interpretation we can define scattering matrix by the formula $S=S_+^*S_-$, however this definition gives a unitary operator only in the case when the image of $S_-$ coincides with the image of $S_+.$

Until now all of our considerations were applicable both to  algebras over real numbers and algebras over complex numbers.  In what follows we restrict ourselves to algebras over complex numbers. Notice, however, that we can apply the considerations below to algebras over real numbers complexifying  the elementary space $\textgoth {h}$ and the pre Hilbert space $\cal H$. 

Let us diagonalize the matrix $E(\bk)$; corresponding eigenvalues  are denoted by $\epsilon_j(\bk)$ and  eigenvectors  are denoted $\rho_j^{\alpha}(\bk).$ ( We assume that these eigenvectors constitute an orthonormal system.)  In momentum representation generalized eigenvectors vectors of time and spatial translations  are $\rho_j^{\alpha}({\bk})\delta ({\bk}-{\bk}_0).$
We consider $in$- and $out$-operators corresponding to these eigenvectors as generalized functions of $\bk$; they are denoted 
$$a_{in}({\bk},j), a_{in}^+({\bk},j),
a_{out}({\bk},j), a_{out}^+({\bk},j).$$
For example, $\int d{\bk}\sum_if^i({\bk}) a_{out}^+({\bk},i)= a_{out}^+(f\rho)$ where $f\rho=f^i({\bk})\rho_i^{\alpha}({\bk})).$ (Here $\rho$ is considered as a square matrix.)
These operators can be interpreted as annihilation and creation $in$- and $out$- operators of particles with momentum $\bk.$
Sometimes we omit discrete indices characterizing the type of particle in these operators, then the operators $a_{in}({\bk}), a_{in}^+({\bk}),
a_{out}({\bk}), a_{out}^+({\bk})) $ should be regarded as $m$-dimensional vectors and  the values of corresponding correlation functions as  elements of tensor product of $m$-dimensional spaces.

 If we  assume that  the theory has particle interpretation, the M\o ller matrices establish unitary equivalence of  $a_{in}({\bk},j), a_{in}^+({\bk},j),
a_{out}({\bk},j), a_{out}^+({\bk},j)$ with $b({\bk},j), b^+({\bk}, j)$ where 
$b({\bk},j), b^+({\bk}, j)$ are operator generalized functions in ${\cal H}_{as}$ corresponding to $b(f\rho),b^+(f\rho).$

The matrix elements of scattering matrix can be expressed in terms of $in$- and $out$-operators:

$$\langle S(b^+(g_1)\cdots b^+(g_n)|0\rangle), 
(b^+(h_1)\cdots b^+(h_m)|0\rangle)\rangle=$$
$$\langle S_-(b^+(g_1)\cdots b^+(g_n)|0\rangle),
S_+(b^+(h_1)\cdots b^+(h_m)|0\rangle) \rangle=$$
$$\langle a_{in}^+(g_1)\cdots a_{in}^+(g_n)\Phi, a_{out}^+(h_1)\cdots a_{out}^+(h_m)\Phi \rangle=$$
$$\langle \lim_{\tau\to-\infty}B(g_1\phi^{-1},\tau)\cdots B(g_n\phi^{-1},\tau)\Phi, \lim _{\tau\to+\infty}B(h_1\phi^{-1},\tau)\cdots B(h_m\phi^{-1},\tau)\Phi\rangle$$
hence

$$\langle S(b^+(g_1)\cdots b^+(g_n)|0\rangle), 
(b^+(h_1)\cdots b^+(h_m)|0\rangle)\rangle=$$
\begin {equation}\label {S}
\lim_{\tau\to-\infty,\tau'\to +\infty} \omega (B^*(h_m\phi^{-1},\tau')\cdots B^*(h_1\phi^{-1},\tau')B(g_1\phi^{-1},\tau)\cdots B(g_n\phi^{-1},\tau))
\end {equation}

and
\begin {equation}\label {SS}
\langle a_{in}^+({\bk}_1,i_1)...a_{in}^+({\bk}_n,i_n)\Phi,a_{out}^+({\bk}'_1,j_1)...a_{out}^+({\bk}'_m,j_m)\Phi\rangle=
\end{equation}
$$ \lim_{\tau\to-\infty,\tau'\to +\infty} \omega (B'({\bk}'_m,j_m,\tau')...B'({\bk}'_1,j_1,\tau') B({\bk}_1,i_1,\tau) ...B({\bk}_n,i_n,\tau))$$
where $$\int d{\bk}\sum_if^i({\bk})B({\bk},i,\tau)=B(f\rho\phi^{-1},\tau),$$
$$ \int d{\bk}\sum_if^i({\bk})B'({\bk},i,\tau')=B^*(f\rho\phi^{-1},\tau').$$

Omitting discrete indices  and using (\ref {BB})  we can write 

$$
\langle a_{in}^+({\bk}_1)...a_{in}^+({\bk}_n)\Phi,a_{out}^+({\bk}'_1)...a_{out}^+({\bk}'_m)\Phi\rangle=$$
\begin {equation}\label {BBB} \lim_{\tau\to-\infty,\tau'\to +\infty} \omega (D({\bk}'_m)B'({\bk}'_m,\tau')...D({\bk}_1)B'({\bk}'_1,\tau') D({\bk}_1)B({\bk}_1,\tau) ..D({\bk}_n)B({\bk}_n,\tau))=\end{equation}
$$ \lim_{\tau\to-\infty,\tau'\to +\infty} \omega (D({\bk}'_m)e^{i\tau'E({\bk}'_m)}\hat B^*({\tau',\bk}'_m)...D({\bk}'_1)e^{i\tau'E({\bk}'_1)}\hat B^*({\tau',\bk}'_1) $$
$$D({\bk}_1)e^{i\tau E({\bk}_1)}\hat B(\tau,{\bk}_1) ..D({\bk}_n)e^{i\tau E({\bk}_n)}\hat B(\tau, {\bk}_n))$$
where $D$ stands for the matrix $\rho \phi^{-1}.$

It is  easy to describe the joint spectrum of momentum and energy operators in ${\cal H}_{as}$ (of infinitesimal generators of spatial and time translations). It consists of
points $( \epsilon_{j_1}({{\bk}_1})+... 
+\epsilon_{j_r}({{\bk}_r}), {\bk}_1+...+{\bk}_r).$ For $r=0$ we obtain the point $(0,0)$ corresponding to the vacuum vector. The points with $r=1$ constitute
one-particle spectrum, the points with $r>1$ belong to multi-particle spectrum.
If the theory has particle interpretation the same formulas describe the joint spectrum of  momentum and energy operators in $\bar {\cal H}.$

The decomposition of the spectrum in one-particle spectrum and multi-particle spectrum induces the decomposition of   $\bar {\cal H}$ into direct sum of one-dimensional space ${\cal H}_0$ containing $\Phi$,
one-particle space ${\cal H}_1$ (the closure of the image of the map $\textgoth {h}\to \cal H$ ) and multi-particle space  $\cal M$.
\section {LSZ formula}
The  scattering matrix can be expressed in terms of Green functions. These functions can be defined by the formula
$$ G_r(\tau _1, {\bx}_1, i_1,...,\tau_r,{\bx}_r,i_r)=
\omega(T(B_{i_1}(\tau _1, {\bx}_1)...B_{i_r}(\tau_r,{\bx}_r))$$
where $T$ stand for the chronological product. We defined Green functions in $(\tau, \bx)$-representation, taking Fourier transform with respect to $\bx$ we  obtain Green
 functions in $(\tau,\bk)$- representation; taking inverse Fourier transform with respect to $\tau$ we get Green functions in $(\varepsilon, \bk)$-representation. For simplicity we are assuming that that $B_i$ are self-adjoint (otherwise we should consider not only $B_i$, but also $B^*_i$ under the sign of $T$-product).

 We have defined Green functions using the operators $B_i$  (good operators), however one can modify this definition replacing $B_i$ by other operators $A_i\in \cal A.$
 
 It is easy to calculate the two-point Green function $G_2.$ We start with two-point correlation function
 $$w_2(\tau_1,{\bx}_1, i,\tau_2,{\bx}_2,j)=\omega (B_i(\tau_1,{\bx}_1)B_j(\tau_2,{\bx_2}))=$$ $$\langle B_j(\tau_2,{\bx_2})\Phi,B_ i(\tau_1,{\bx}_1)\Phi \rangle=\int d{\bk} e^{i{\bk}({\bx_2-\bx_1})} \langle e^{-i\tau_2E({\bk})}\phi_j({\bk}), e^{-i\tau_1E({\bk})}\phi_i({\bk})\rangle$$

 Expressing $G_2$ in terms of $w_2$  we obtain that in $(\varepsilon,\bk)$-representation 
$$G_2(\varepsilon_1,{\bk}_1,i,\varepsilon_2,{\bk}_2,j)=G^{i,j}(\varepsilon_1,{\bk}_1)\delta(\varepsilon_1+\varepsilon_2)\delta ({\bk}_1+{\bk}_2)$$
 where for fixed $\bk$ the function $G^{i,j}(\varepsilon,\bk)$ has poles with respect to $\varepsilon$ at the points $\pm \epsilon_s(\bk)$  (here $\epsilon_s(\bk)$ are eigenvalues of the matrix $E(\bk)$)).   Namely, $$G^{i,j}(\varepsilon,{\bk})= A^{i,j}(\varepsilon,{\bk})+A^{j,i}(-\varepsilon,-\bk)$$
 where
 \begin{equation}\label {AA}
 A^{i,j}(\varepsilon,{\bk})=\sum_s \frac {i}{\varepsilon+\epsilon_s ({\bk})-i0}\langle a_s({\bk})\phi_j({\bk}),\phi_i({\bk})\rangle
 \end{equation}
 We have used the fact that the matrix $e^{iE({\bk})\tau}$ can be expressed as a linear combination of exponents $e^{ i\epsilon_j({\bk})\tau}$ with matrix coefficients depending on $\bk$:
 $$e^{iE({\bk})\tau} =\sum _s a_s({\bk}) e^{i\epsilon_s({\bk})\tau}. $$

 Using the same fact and (\ref {S})  or (\ref {BBB}) it is easy to check that  the scattering matrix can be expressed in terms of asymptotic behavior of Green functions in $(\tau, \bk)$ representation. (One should divide the arguments of Green function in two groups; in one group we should take the times tending to $-\infty$, in the second group to $+\infty$. The ordering of times in every group is irrelevant due to asymptotic commutativity of factors in (\ref {S}). )
 
 Equivalently one can work in $(\varepsilon, \bk)$- representation taking inverse Fourier transform with respect to $\tau$ in $(\tau,\bk)$-representation. Then the scattering matrix can be expressed in terms of poles of Green functions with respect to $\varepsilon$ and residues in these poles.  This is the famous LSZ formula. One can derive it from the following statements:
 
 {\it  Let  $E$ denote a Hermitian matrix with eigenvalues $\epsilon_j$. Then the matrix $e^{i\tau E}$ can be written in the form $\sum_j a_je^{i\tau \epsilon_j}$
 where $a_j$ are constant matrices.
 
  Let us assume  there exist limits $A_{\pm}=\lim_{\tau\to \pm \infty} e^{i\tau E}\rho(\tau)$ where $\rho$ is a column vector. Then $\rho(\tau)$ has asymptotic behavior 
  $$\rho(\tau)\sim   e^{-i\tau E}A_{\pm}= \sum_k e^{-i\epsilon_k\tau}a_{k}A_{\pm}$$
 as $\tau\to\pm\infty.$ 
 
 This implies that the (inverse) Fourier transform $\rho(\varepsilon)$ of $\rho(\tau)$ has poles at the points
 $\pm (\epsilon_k +i0)$ with residues  $\mp2\pi i a_k A_{\pm}.$}
 
 We can say that
 
 {\it The asymptotic behavior of $\rho (\tau)$ is determined by the polar part of $\rho (\varepsilon).$}
 
In what follows we use these statements in a little bit different form.
We  represent $\rho(\tau)$   as $ e^{-i\tau E}A_{-}\Phi(-\tau)+e^{-i\tau E}A_{+}\Phi(\tau) +\sigma (\tau)$  where $\sigma(\tau)\to 0$ as $\tau\to\infty.$ We assume that $\sigma(\tau)$ is a summable function, then
$$\rho (\varepsilon)=(E-\varepsilon+i0)^{-1} C_- +(E+\varepsilon+i0)^{-1} C_++\sigma (\varepsilon)$$
where $\sigma (\varepsilon)$ is continuous, $C_-$ and  $C_+$ do not depend on $\varepsilon .$ We say that the first two summands constitute the polar part of  $\rho (\varepsilon).$  We need the following statement

{\it If $f(\varepsilon)$ is a smooth function then 
\begin {equation} \label {POL}
 f(\varepsilon)\rho (\varepsilon)=(E-\varepsilon-i0)^{-1} C'_- +(E+\varepsilon-i0)^{-1} C'_+\sigma '(\varepsilon)
 \end{equation}
 where $C'_-=f(E)C_-, C'_+=f(-E)C_+$ do not depend on $\varepsilon$ and $\sigma'$ is a continuous function.}
 
 Notice that in the LSZ formula the operators $B_i$ transforming the vector $\Phi$ into  an element of $\Phi (\textgoth {h})$ (of  one-particle space) can be replaced by asymptotically   commuting smooth  operators $A_i\in \cal A$ obeying a weaker condition: the projections of vectors $A_i\Phi$ on the one-particle space are linearly independent and the projection  of $A_i\Phi$ on $\Phi$ vanishes. This can be proved if  the theory has particle interpretation. ( See Section 4.6 of \cite {SC} for the proof of this fact  in less general case; this proof can be generalized to our setting.)  Instead  of this condition we can require the existence of smooth fast decreasing functions $\alpha_{i}^j(\tau, \bx)$ such that the operators $B_i=\int d\tau d {\bx}\alpha_i^j(\tau,{\bx})A_j(\tau,\bx)$ are good operators.)  ( This condition is always satisfied if   the joint spectrum of $H,\bP$ in ${\cH}_0\oplus {\cH}_1$
  does not overlap with multi-particle spectrum.) 
  
  Using the formula
 $$ B_i(\tau,{\bx})=\int d\tau' d{\bx}'\alpha_i^j (\tau'-\tau, {\bx}'-{\bx}) A_j(\tau',{\bx}') $$
 we can  express the correlation functions for operators $B_i$ in terms of correlation functions for operators $A_i$. The expression   looks very simple in $(\varepsilon, {\bk})$-representation: for example if $\alpha _i^j=\alpha_i\delta_i^j$ one should multiply the correlation functions of operators $A_i$ by the product of Fourier transforms of functions $\alpha_i$  with respect to $\tau,\bx.$ The corresponding expression for Green functions is more complicated due to factors
 $\theta(\tau_i-\tau_j)$ entering the definition of chronological product. However, in scattering theory we are interested in   asymptotic behavior of Green functions in $(\tau, \bk)$ representation or in the behavior of  polar parts of Green functions in $(\varepsilon,\bk)$-representation. For $\bk$ in dense open set the behavior of the polar parts of Green functions for operators $A_i$ in $(\varepsilon,\bk)$ representation  can be described in the same way as for correlation functions. ( To  prove this statement we use asymptotic commutativity of operators $A_i$ and the assumption (\ref {N}). In the calculation of scattering matrix we decompose the arguments of Green functions  in two groups; we use the fact that due to asymptotic commutativity the time ordering inside every group is irrelevant.)
 
 Let us give more precise formulations of the above statements.
 
 We are starting with asymptotically commuting operators $A_1,...,A_m$  obeying $\langle A_i\Phi,\Phi \rangle=0.$
  We introduce the notation  $T^A_i ({\bk})$ for projections of vectors $A_i\Phi\in \cal H$ on $\Phi (\textgoth {h})$ (on one-particle subspace of $\cal H$) considered as  elements of $\textgoth {h}$ in momentum representation. We assume that these projections are linearly independent (the matrix $T^A({\bk})=(T^A({\bk}))_i^{\alpha}$ is non-degenerate). 
 
 We consider Green function
 $$ G_r(\tau _1, {\bx}_1, i_1,...,\tau_r,{\bx}_r,i_r)=
\omega(T(A_{i_1}(\tau _1, {\bx}_1)...A_{i_r}(\tau_r,{\bx}_r))$$
and their Fourier transforms (Green functions in $(\tau, \bk)$- and $(\varepsilon,\bk)$- representations. Notice that due to translation invariance the Green function in  $(\varepsilon,\bk)$- representation contains a delta-function $\delta(\sum \varepsilon_i)\delta(\sum {\bk}_i)$; talking about two-point function (r=2) we always omit  this delta-function. (Hence the two-point Green function is a matrix-valued function of $(\varepsilon,\bk)$.)
We can write the two-point Green function in $(\varepsilon,\bk)$-representation  as a sum of the polar part  (having first order poles with respect to variables $\varepsilon$)  and regular part. 
The polar part 
$$(E({\bk})-\varepsilon+i0)^{-1} C_-({\bk})+(E({\bk})+\varepsilon+i0)^{-1} C_+({\bk})$$
 governs the behavior of Green function in $(\tau, {\bk})$-representation as $\tau\to\infty$; it  is a sum of two summands; one of them  ($in$-summand) is responsible for the limit $\tau\to -\infty$, another ($out$-summand) is responsible for the limit $\tau\to\infty$.

Let us consider the Green function $G_r$ in $(\tau, \bk)$-representation. We assume that the arguments of this function are divided in two groups  (with indices $i$  in the interval $1\leq i\leq a$ and with indices in the interval $a<i\leq r$.) We assume that the times $\tau_i$ with the indices from the first group tend to $-\infty$ and the remaining times tend to $+\infty.$

In  $(\varepsilon,\bk)$- representation the Green function $G_r$ can be represented as a product of the amputated Green function   and $r$  two-point Green functions labeled by index $i.$ We change the signs of the variables $\varepsilon_i, {\bk}_i$ where $i$ is the index from the second group to interpret these variables as energies and momenta of outgoing particles.  We define the polar part 
$$ P_r(\varepsilon_{1},{\bk}_1,j_1,...,\varepsilon_{r},{\bk}_r,j_r)$$
 of  Green function replacing every two-point Green function in this representation by its $in$-summand of its polar part for indices $i\leq a$ and by $out$-summand of the  polar part for $i>a$.

  Let us define operators $A'_i$ by the formula
$A'_i=\int d\tau d{\bx} \alpha_i^j(\tau,{\bx})A_j(\tau,\bx)$ where $\alpha_i^j(\tau,{\bx})$ are smooth fast decreasing functions. Polar parts of corresponding  Green functions are denoted by by $P'_r.$

It is easy to express the projections $T^{A'}_i({\bk})$ of $A'_i\Phi$ on one-particle space in terms of projections
$T^A_j({\bk}).$ We obtain
\begin {equation}
\label{AA}
T^{A'}_i({\bk})=\alpha_i^j (E({\bk}),{\bk}) T^A_j({\bk})
\end {equation}
where
$\alpha_i^j (E ({\bk}),{\bk})=\int d\tau d {\bx}e^{i{\bk \bx}-iE({\bk})\tau}\alpha_i^j(\tau,{\bx})$

Equivalently
\begin{equation}\label {TT}
T^{A'}({\bk})=\alpha(E({\bk}),{\bk})T^A({\bk}).
\end{equation}
To prove (\ref {AA})  we represent $A_j(\tau,{\bx})\Phi$ as $\Phi (T^{A(\tau,{\bx}}_j)+\rho(\tau,{\bx})$ where $\rho$ belongs to the multiparticle space. We get
$$ A'_i\Phi=\int d\tau d{\bx} \alpha_i^j(\tau,{\bx})\Phi (T^{A(\tau,{\bx}}_j)+\int d\tau d{\bx} \alpha_i^j(\tau,{\bx})\rho (\tau,{\bx}).$$
Noticing that the second summand lies in the multiparticle space and that for $\phi \in \textgoth {h}$ we have
$\Phi(e^{i{\bP \bx}-H\tau}\phi)=\Phi( e^{i{\bk\bx}-iE({\bk})\tau}  \phi)$ we obtain (\ref {AA}). 

As we noticed the polar part governs the asymptotic behavior in $(\tau,\bk)$- representation. Using this fact one can prove that $P'_r(\varepsilon_{1},{\bk}_1,i_1,...,\varepsilon_{},{\bk}_r,i_r)$ is equal to the polar part of

\begin {equation}\label {PO}
\alpha_{i_1}^{j_1}(\varepsilon_{1},{\bk}_1)...\alpha_{i_r}^{j_r}(\varepsilon_{r},{\bk}_r)P_r(\varepsilon_{1},{\bk}_1,j_1,...,\varepsilon_{r},{\bk}_r,j_r).
\end {equation}
Using (\ref {POL}) we obtain that
$$P'_r(\varepsilon_{1},{\bk}_1,i_1...,\varepsilon_{},{\bk}_r,i_r)= \alpha_{i_1}^{j_1}(E({\bk}_1),{\bk}_1)...\alpha_{i_r}^{j_r}(E({\bk}_r),{\bk}_r)P_r(\varepsilon_{1},{\bk}_1,j_1,...,\varepsilon_{r},{\bk}_r,j_r)$$

or equivalently
\begin {equation}\label {PT}
P'_r(\varepsilon_{1},{\bk}_1,...,\varepsilon_{},{\bk}_r)= \alpha(E({\bk}_1),{\bk}_1)\otimes ...\otimes \alpha(E({\bk}_r),{\bk}_r)P_r(\varepsilon_{1},{\bk}_1,...,\varepsilon_{r},{\bk}_r)
\end{equation}

In this formula and in what follows  we consider $P'_r$ and $P_r$ as functions taking values in $r$-th tensor power of $m$-dimensional space 
(we consider discrete variables in $P_r$ and in $P'_r$ as tensor indices).

Let us define now the normalized polar part of Green function (closely related to  Green function on shell) 
by the formula
$$ \tilde P_r(\varepsilon_{1},{\bk}_1,...,\varepsilon_{r},{\bk}_r)
=(T^A({\bk}_1))^{-1}\otimes ...\otimes (T^A ({\bk_r}))^{-1}P_r(\varepsilon_{1},{\bk}_1,...,\varepsilon_{r},{\bk}_r).$$
This function takes values in $r$-th tensor power of $\textgoth {h}$.

It follows immediately from (\ref {TT}), (\ref {PT})   that 
\begin {equation}\label {NT}
\tilde P'_r(\varepsilon_{1},{\bk}_1...,\varepsilon_{r},{\bk}_r)=\tilde P_r(\varepsilon_{1},{\bk}_1...,\varepsilon_{r},{\bk}_r).
\end {equation}
This means that normalized polar parts of Green functions for $A_1,...,A_m$ and $A'_1,...,A'_m$ coincide.

Let us define Green functions on shell taking the  residues of normalized polar parts of Green functions.

 Now we can formulate the LSZ formula in the following way:
 
 {\it  Matrix elements of scattering matrix  coincide with Green functions on shell (up to sign change in outgoing momenta)}
 
 To prove this fact it is sufficient to  verify it for good operators $B_1,...,B_m$ using  (\ref {S}) and (\ref {BBB}).
 
 Notice that the matrices $T^A$ entering the definition of normalized polar part are closely related to  the polar part of two-point Green function.
 \section {Inclusive scattering matrix}
 An element $B$ of the algebra $\cal A$ specifies two operators on linear functionals on $\cal A$. First operator is denoted by the same symbol $B$; it transforms the functional  $\sigma$ into the functional $(B\sigma)(A)=\sigma(AB).$ The second one (denoted $\tilde B$) transforms $\sigma$ into the functional
 $(\tilde B\sigma )(A)=\sigma (B^*A).$
The vectors in the space $\cal H$ correspond to excitations of the state $\omega$:  the vector $B\Phi$ corresponds to the state $\tilde B B\omega.$ 
Let us introduce  notations
$$ B(f)=B(f,0), L(g)=\tilde B(g\phi^{-1})B(g\phi^{-1}),  L(g,\tau)=T_{\tau}L(T_{-\tau}g)=\tilde B(f,\tau) B(f,\tau)$$
where $g=f\phi\in \textgoth {h}.$
Instead of working with vectors 
$$\Psi (f_1,\tau_1, ..., f_n,\tau_n)= B(f_1, \tau_1) ...B(f_n,\tau_n)\Phi$$
we can work with corresponding states
$$ \Lambda (g_1,\tau_1,...,g_n,\tau_n)=L(g_1,\tau _1),...L(g_n,\tau_n)\omega.$$ It follows from Lemma \ref {NO} that these states have a limit as $\tau_j\to -\infty$  if $f_j=g_j\phi^{-1}$ do not overlap. These states  will be necessary in geometric approach (see \cite {GA3}), however they are
useful also in algebraic approach. Namely, we use these states to construct the inclusive scattering matrix.

Let us consider the state 
$$L(g'_1,\tau' _1)...L(g'_{n'},\tau'_{n'})L(g_1,\tau _1)...L(g_n,\tau_n)\omega$$
considered as a linear functional on $\cal A$ (as an element of the cone $\cal C$).
We assume that $g'_i$ as well as $g_j$ are not overlapping, then 
 this state has a limit as $\tau '_i\to +\infty, \tau_j\to -\infty$; we denote this limit by $Q$. Notice that $Q$ does not change if we permute $g_1, ...,g_n$ (in the limit $\tau_j \to -\infty$ the operators $L(g_j,\tau_j)$ commute). Similarly $Q$ does not change if we permute $g'_1,...,g'_{n'}.$

More generally we can consider a  linear functional $\tilde Q$ on $\cal A$ defined as a limit of
$$L(\tilde g'_1,g'_1,\tau' _1)...L(\tilde g'_{n'}, g'_{n'},\tau'_{n'})L(\tilde g_1,g_1,\tau _1)...L(\tilde g_n,g_n,\tau_n)\omega$$  as $\tau '_i\to +\infty, \tau_j\to -\infty$.( We introduced the notation  $L(\tilde g, g, \tau)=\tilde B(\tilde f,\tau) B(f,\tau)$.)

Then we define 
$$\sigma (\tilde g'_1,g'_1,..., \tilde g'_{n'}, g'_{n'},\tilde g_1,g_1,...,\tilde g_n, g_n).$$
as $\tilde Q(1).$ This functional is linear with respect to its arguments $\tilde g'_i, g_i, \tilde g_j, g_j.$ It is well defined if  each of four families $\tilde g'_i, g_i, \tilde g_j, g_j$ is non-overlapping. In bra-ket  notations

$$\sigma (\tilde g'_1,g'_1,..., \tilde g'_{n'}, g'_{n'},\tilde g_1,g_1,...,\tilde g_n, g_n)=$$
\begin {equation} \label {BK}
\langle 1|  \lim _{\tau'_i\to +\infty, \tau_j\to -\infty} L(\tilde g'_1,g'_1,\tau' _1)...L(\tilde g'_{n'}, g'_{n'},\tau'_{n'})L(\tilde g_1,g_1,\tau _1)...L(\tilde g_n,g_n,\tau_n)|\omega\rangle
\end {equation}

By definition the functional  $\sigma$ is  {\it inclusive scattering matrix}.

To justify this definition we notice that
$$\sigma (\tilde g'_1,g'_1,..., \tilde g'_{n'}, g'_{n'},\tilde g_1,g_1,...,\tilde g_n, g_n)=$$
$$ \lim_{\tau_j\to -\infty',\tau'_i\to +\infty}(L(\tilde g_1,g_1,\tau _1)...L(\tilde g_n,g_n,\tau_n)\omega)
(B(f'_1,\tau'_1)... B(f'_{n'},\tau'_{n'})B(f'_{n'},\tau'_{n'})^*...B(f'_1,\tau'_1)^*)=$$
$$( \lim_{\tau_j\to -\infty}(L(\tilde g_1,g_1,\tau _1)...L(\tilde g_n,g_n,\tau_n)\omega)
(a^+_{out}(g'_1)...a^+_{out}(g'_{n'})a_{out}(\tilde g'_{n'})...(a_{out}(\tilde g'_1))$$
We have used the relation $(\tilde M N\rho)(X)=\rho (M^*XN)$ in this derivation.

The inclusive cross-section can be expressed  in terms of inclusive $S$-matrix defined above. To verify this statement we
consider the expectation value
\begin{equation}
 \label{eq:NU}
 \nu (a^+_{out, k_1}({\bp}_1)a_{out, k_1}({\bp}_1)\dots a^+_{out,k_m}({\bp}_m)a_{out,k_m}({\bp}_m))
\end{equation}
where $\nu $ is an arbitrary state.
This quantity is the probability density in momentum space for finding $m$ outgoing particles of the types $k_1,\dots ,k_n$ with momenta $\bp_1,\dots ,\bp_m$ plus other unspecified outgoing particles.
It gives inclusive cross-section if $$\nu= \lim _{\tau _i\to -\infty}L(g_1,\tau _1)...L(g_n,\tau_n)\omega$$

The inclusive scattering matrix can be expressed in terms of generalized Green functions (GGreen functions). These functions appear naturally in the formalism of $L$-functionals  \cite {TS}, \cite {T},
\cite {MO}; their relation to  inclusive cross-sections is analyzed in \cite {S}, \cite {SC}, \cite {MO}. They appear also in Keldysh formalism and in thermo-field dynamics  \cite{UNI},\cite {CU},\cite {K}. GGreen functions can be defined by the formula
$$ G (\tau _1, {\bx}_1, i_1,...,\tau_r,{\bx}_r,i_r,\tau'_1,{\bx}'_1,i'_1,...,\tau'_{r'},{\bx}'_{r'},i'_{r'})=$$
$$(T(B_{i_1}(\tau _1, {\bx}_1) ...B_{i_r}(\tau_r,{\bx}_r,)\tilde B_{i'_1}(\tau'_1,{\bx}'_1)...\tilde B_{i'_{r'}}(\tau'_r,{\bx}'_r))\omega)(1)$$
where $T$ stand for the chronological product. More precisely we defined GGreen functions in $(\tau, \bx)$-representation, taking Fourier transform with respect to $\bx$ we  obtain GGreen
 functions in $(\tau,\bk)$- representation.  Using the fact that the matrix $e^{iE({\bk})\tau}$ can be expressed as a linear combination of exponents $e^{ i\epsilon_j({\bk})\tau}$ with matrix coefficients depending on $\bk$ it is easy to check that  the inclusive scattering matrix can be expressed in terms of asymptotic behavior of GGreen functions in $(\tau, \bk)$ representation. (One should take $r=r'$ and assume that $\tau_i\to +\infty,\tau'_i\to+\infty$ for $i\leq m$ and $\tau_j\to-\infty,\tau'_j\to -\infty$ for $j>m$. The ordering of times in every group is irrelevant due to asymptotic commutativity of factors. )

 Equivalently one can work in $(\varepsilon, \bk)$- representation taking inverse Fourier transform with respect to $\tau$ in $(\tau,\bk)$-representation. Then the inclusive scattering matrix can be expressed in terms of poles of GGreen functions with respect to $\varepsilon$ and residues in these poles. 
 
 As in LSZ formula for scattering matrix we can work with operators $A_1,...,A_m$ requiring the existence of fast decreasing functions $\alpha _i^j$ such that the operators $B_i=\int d\tau d{\bx} \alpha_i^j(\tau,{\bx})A_j(\tau,\bx)$ are good operators. Using K$\rm \dot {a}$ll$ \acute{e}$n-Lehmann representation of two-point GGreen function we define polar part and normalized polar part of GGreen function. (We represent GGreen functions in terms of amputated GGreen functions and use 
  K$\rm \dot {a}$ll$ \acute{e}$n-Lehmann representation of two-point GGreen function in the proof.) 
  
 \section {Fermions}
 We assumed that operators $B_i$ asymptotically commute (\ref {AS}).
 One can replace this assumption with the assumption of asymptotic anticommutativity  (we replace the commutator in (\ref {AS}) by anticommutator).
 Then we should modify also the definition of truncated correlation functions including some signs.
 
 One can repeat all considerations of present paper in this situation. Slight modifications are necessary. In particular, instead of bosonic Fock space one should consider fermionic Fock space (Fock representation of canonical anticommutation relations.) The particles obey Fermi statistics.
 To define inclusive scattering matrix we again consider states
 $$ \Lambda (g_1,\tau_1,...,g_n,\tau_n)=L(g_1,\tau _1),...L(g_n,\tau_n)\omega$$
 and prove that these states have a limit as $\tau_i\to -\infty$ under the same conditions on $g_1, ...,g_n.$  It is important to notice that under these conditions it follows from the asymptotic {\it anticommutativity} of operators $B_i$ that the operators  $L(g_i, \tau_i)$ {\it commute} in the limit  $\tau_i\to -\infty.$
 \section {BRST formalism}
 Methods of homological algebra (=BRST formalism)  can be applied in scattering theory.
 Recall that in homological algebra together with  modules (algebras, etc) one considers differential graded modules (algebras,...).  It is sufficient to have $\mathbb{Z}_2$-grading. A module is $\mathbb{Z}_2$-graded if it is represented as a direct sum of even and odd parts.
  A differential  can  be defined as parity reversing homomorphism $Q$ obeying $Q^2=0.$ Homology is defined as $Ker Q/Im Q$ (as a quotient of the submodule consisting of $Q$-closed  elements with respect to the submodule consisting of $Q$-exact elements).  
  
  The main idea is to replace a module  by simpler (for example, free) differential graded module. (The new module should be quasi-isomorphic to the original module, considered as a differential module with trivial grading and trivial differential.  Quasi-isomorphism is defined as a homomorphism
commuting with the differential  and inducing an isomorphism on homology.)

The above considerations can be applied to differential  $\mathbb{Z}_2$- graded algebras (algebras with parity reversing BRST  operator $Q$  obeying $Q^2=0.$) All physical quantities should be BRST-closed (should belong to the kernel of $Q$); one should neglect the BRST-exact  quantities  (the elements of the image of $Q$). The BRST-operator on algebra should satisfy graded  Leibniz rule: $Q(AB)=Q(A)B\pm AQ(B)$ (plus sign if $A$ is even, minus sign if $A$ is odd. If algebra $\cal A$  is realized by operators in differential $\cal A$-module with  differential $\hat Q$ then the differential $Q$ on algebra is defined as supercommutator with  $\hat Q$,
 i.e. $QA=[Q,A]$ if $A$ is even and $QA=[Q,A]_+$ if  $A$ is odd. 
 
 Instead of Hilbert spaces one can consider differential modules equipped with a structure of pseudo Hilbert space (space with non-degenerate, but indefinite scalar product). However, the indefinite scalar product should descend to definite scalar product on homology. 
 
   Gelfand-Naimark-Segal (GNS) construction can be generalized to the case when an algebra is not equipped with involution. In this generalization we start with 
  a unital associative algebra $\cal A$ and a linear functional $\omega$ on $\cal A.$ Then we can introduce a (not necessarily symmetric) scalar product on $\cal A$ by the formula $\langle x,y\rangle=\omega (xy$). We are saying that $a\in \cal A$ is a right null vector if $\langle x,a\rangle=0 $ for every $x\in \cal A.$ It is easy to check that right null vectors constitute a left ideal in $\cal A$.  Factorizing $\cal A$  with respect to this ideal we obtain a right $\cal A$-module denoted by $R.$ Similarly factorizing $\cal $  with respect to  left null vectors we obtain a left $\cal A$-module denoted by $L.$
 It is easy to define a pairing between $L$ and $R$; this paring is non-degenerate. If the algebra $\cal A$ is equipped with involution  we can consider an induced involution on the space of linear functionals; we assume that the functional $\omega$ is self-adjoint. Then $R$ is complex conjugate  to $L$ and the pairing between $L$ and $R$ can be interpreted as  (in general indefinite) scalar product in $L$. If $\omega$ is a positive functional we come back to the
  GNS construction.
 
  Let us suppose now $\cal A$ is a differential algebra with differential $Q$. This differential specifies a differential on the space ${\cal A}^{\vee}$ of linear functionals denoted by the same symbol. {\it We assume that
  $Q\omega=0$ (the functional $\omega$ is $Q$-closed).}  This assumption implies that the ideals we constructed are $Q$-invariant, hence the differential $Q$ descends to  the $\cal A$-modules $R$ and $L$. The pairing between differential modules $R$ and $L$ respects the differential $Q.$
  
  We will work  with differential algebra $\cal A$ equipped with involution $^*$ that agrees with the differential.
  We  assume that time translations and spatial translations act as automorphisms of $\cal A$ and commute with the differential $Q.$  We fix a translation-invariant stationary self-adjoint  $Q$-closed linear functional $\omega$ that descends to a positive functional on homology of $\cal A$.
  Applying the modification of GNS construction to $\omega$ we obtain a
  differential pseudo pre Hilbert space $\cal H$ with the differential (BRST operator) denoted $\hat Q$.
  
  We modify the definition of elementary space  saying that a differential vector space $\tilde{\textgoth{h}}$ is an elementary space in new sense if its homology can be identified with elementary space in old sense.  An elementary excitation of $\omega$  is defined as an  linear map of
$\tilde{\textgoth{h}}$ in $\cal H$ commuting with space-time translations and differentials (BRST operators).
 
 We can repeat with minor modifications  the construction of M\o ller matrices and scattering matrices in  new situation. In particular, a scattering matrix $\tilde S$ can be defined as an operator in Fock space $\tilde F$ corresponding to the space $\tilde{\textgoth{h}}$. The operator $\tilde S$ commutes with BRST operator, hence it descends to homology giving the scattering matrix $S$ of physical (quasi)particles. (The operator $S$ acts in the Fock space $\tilde F$ corresponding to the space ${\textgoth{h}}.$)


 {\bf Acknowledgements} I am indebted to A. Kapustin for valuable discussions.
 

\begin{thebibliography}{10}
 \bibitem {KA} Kapustin, A., 2013. Is there life beyond Quantum Mechanics?. arXiv preprint arXiv:1303.6917.
\bibitem {GA1} Schwarz A. Geometric approach to quantum theory. SIGMA. Symmetry, Integrability and Geometry: Methods and Applications. 2020 Apr 1;16:020.
\bibitem {GA} Schwarz A.  2021. Geometric and algebraic approaches to quantum theory. arXiv preprint arXiv:2102.09176.
 Nuclear Physics B, 973, p.115601.
\bibitem {GA3}Schwarz, A., 2021. Scattering in geometric approach to quantum theory. arXiv preprint arXiv:2107.08557.

\bibitem {GA4} Schwarz A. Scattering  in algebraic approach to quantum theory. Jordan algebras (in preparation)

\bibitem {SCH} A.S. Shvarts, New formulation of quantum theory, Dokl. Akad. Nauk SSSR, 173, 793 (1967).
\bibitem {S} Schwarz A. Inclusive scattering matrix and scattering of quasiparticles. Nuclear Physics B. 2020 Jan 1;950:114869.
\bibitem {SC} Schwarz, A., 2019. Scattering matrix and inclusive scattering matrix in algebraic quantum field theory,
arXiv preprint arXiv:1908.09388

\bibitem {TS} Schwarz, A and Tyupkin, Yu. , On adiabatic definition of S-matrix in the formalism of L-functionals, International seminar on Functional methods in Quantum Field Theory and Statistics, 1971, Lebedev Institute, 38-42

\bibitem{T} Tyupkin, Yu, On the adiabatic definition of the S matrix in the formalism of L-functionals,  Theoretical and Mathematical Physics, 1973, 16:2, 751-756, https://link.springer.com/content/pdf/10.1007\%2FBF01037126.pdf
\bibitem {UNI}Chou, K.C., Su, Z.B., Hao, B.L. and Yu, L., 1985. Equilibrium and nonequilibrium formalisms made unified. Physics Reports, 118(1-2), pp.1-131.
\bibitem {CU} Chu, H., and H. Umezawa. A unified formalism of thermal quantum field theory. International Journal of Modern Physics A 9.14 (1994): 2363-2409.
\bibitem {K}   Kamenev, Alex, and Alex Levchenko. Keldysh technique and non-linear  sigma-model: basic principles and applications. Advances in Physics (2009).
\bibitem {MO} A. Schwarz, Mathematical foundations of quantum field theory, World Scientific 
\bibitem {AH}Araki H, Haag R. Collision cross sections in terms of local observables. Communications in Mathematical Physics. 1967 Apr;4(2):77-91.
\bibitem {FS} Fateev, V.A. and Schwarz, A.S.,  On axiomatic scattering theory. Teoreticheskaya i Matematicheskaya Fizika, 1973, 14(2), pp.152-169.


\end{thebibliography}
\end {document}